\documentclass {article}

\usepackage [utf8] {inputenc}
\usepackage {graphicx}
\usepackage {amsfonts}
\usepackage {amsthm}
\usepackage {amsmath}
\usepackage {subfigure}

\usepackage [english]{babel}
\usepackage [autostyle,english = american]{csquotes}
\MakeOuterQuote{"}

\usepackage{ulem} 

\title{Bayesian statistical analysis to explore the use of glucometer measurements of capillary blood sugar for performing OGTTs}
\date{}
\author{Nicolás E. Kuschinski, J. Andrés Christen, Adriana Monroy\\
CIMAT, Guanajuato, Gto., Mexico}
\begin{document}
\maketitle{}

\begin{abstract}
A common test for the diagnosis of type 2 diabetes is the Oral Glucose Tolerance Test (OGTT). Recent developments in the study of OGTT tests have framed it as a Bayesian inverse problem. These data analysis advances promise great improvements in the descriptive power of OGTTs. OGTT tests are typically done with invasive, bothersome, and somewhat expensive venous blood tests. A natural question is whether improved data analysis techniques would allow for less invasive and cheaper glucometer measurements to be used. In this paper we explore this question. Using one dynamic model, we develop an error model for glucometer capillary blood sugar measurements and compare results of venous blood sugar tests for 65 patients, finding a match in over 90\% of observed cases. Our conclusion suggests that this model (or one much like it) may permit capillary glucose to be used with reasonable accuracy in performing OGTTs.
\end{abstract}

\section{Introduction}

Diabetes is a serious illness that affects a large and growing number of people each year \cite{diabetesprevalence}. According to the World Health Organization, diabetes occurs when the pancreas does not produce enough insulin, or if the insulin is not used effectively \cite{who}.

There are two types of diabetes: Type 1, which is related to genetic disorders and typically diagnosed very early in life, and type 2, which develops over time and often produces few symptoms at the start. With a timely diagnosis and treatment, type 2 diabetes can be relatively easily handled, although it is still a nuisance. If left untreated for too long, however, type 2 diabetes can be a serious health risk and may even be fatal \cite{diabetestreatment}.

In order to diagnose diabetes, a common test is the Oral Glucose Tolerance Test (OGTT). For this test, a fasting patient has his/her blood glucose measured and then drinks a sugary syrup (75g of glucose). Over the course of the next two hours, the patient's blood glucose is measured again at several times, providing information as to how the patient processes blood sugar over time \cite{diabetestreatment, earlymodelogtt, ogttcriteria}.

Classical diagnosis using OGTT results has been very crude, basing results on simple descriptive statistics such as the maximum measurement, the average of measurements, or the value of the last measurement \cite{ogttcriteria}. A somewhat obvious issue with these methods of analysis is that they all ignore that OGTT measurements are measurements of a process over time and that they have a clear correlation structure. Recently, however, mathematical models have been proposed which use a dynamical system to model the patient's blood glucose over the course of the OGTT. Blood glucose variation is described with a system of Ordinary Differential Equations (ODEs), and then a statistical model describes the measurements. The OGTT test has been framed as an inverse problem suitable for Bayesian inference. Although the use of mathematical models for studying OGTTs is not a new development \cite{earlymodelogtt}, they were not seriously used for inference until much more recently. Models have been recently proposed by \cite{odemodel} and \cite{hugo} and have been used to infer some OGTTs with promising results.

One issue with OGTT tests is that they are somewhat expensive and cumbersome. A patient must have a cannula in his/her arm for the duration of the test and results are processed via a slow procedure in a laboratory. It is possible to measure blood glucose cheaply, easily and quickly, using a glucometer, but glucometers have been thought of as an inadequate method that does not have enough precision for an OGTT \cite{glucometererrors}. The improvements in our ability to analyze OGTT results, however, suggest that it may be time to revisit the question of whether glucometer OGTT analysis is possible.

In this paper we first present the ODE model we will be using, and the method by which it is used for inference. Next, we propose a model for error distributions for glucometer measurements and attempt to adapt our model to these new circumstances. This new model is tested with promising results, leading to an improved model which produces satisfactory results in over 90\% of patients.

\section{The ODE Model}

The developments in this paper all relate to the modeling of measurement errors, and should be -- for the most part -- independent of the specific dynamic model considered, so long as it can fit data reasonably. That said, it is necessary to use some kind of dynamic model in order to perform tests, since otherwise there is no way to evaluate the impact on inference. While we expect that the results in this paper behave reasonably for multiple dynamic models, we recommend performing some form of validation if a different dynamic model is used. One dynamic model designed to fit OGTT data is described in detail in \cite{hugo}, and another, closely related model, which we use herein, is described in \cite{odemodel}. This model has been applied in numerous cases with real patients. We proceed to summarize the information about the model. The dynamic model is the following:

\begin{align}
    \frac{dG}{dt} &= L-I+\frac{D}{\theta_2}\label{eq:dyn1}\\
    \frac{dI}{dt} &= \theta_0(G-G_b)^+-\frac{I}{a}\label{eq:dyn2}\\
    \frac{dL}{dt} &= \theta_1(G_b-G)^+-\frac{L}{b}\label{eq:dyn3}\\
    \frac{dD}{dt} &= -\frac{D}{\theta_2}+\frac{2V}{c}\label{eq:dyn4}\\
    \frac{dV}{dt} &= -\frac{2V}{c}\label{eq:dyn5}.
\end{align}

The model represents blood glucose with $G$, in $mg/dl$. $I$ represents the effect of the patient's blood insulin. $L$ represents the response of the patient's body to increase blood sugar if it drops too low (this happens by way of several mechanisms, for example glucagon). $D$ is the glucose in the digestive system and $V$ is the glucose which has not yet been ingested. The values of $a,b,c$ and $G_b$ are known constants taken from literature, and $\theta_0,\theta_1,\theta_2$ and $G(0)$ are the inference parameters.

This model follows a simple three compartment and double feedback logic, as we explain next. As the patient consumes the glucose concentrate, glucose moves into the digestive system (see (\ref{eq:dyn5}) and (\ref{eq:dyn4})). Once in the digestive system, glucose enters the blood at a rate determined by $\theta_2$ (see (\ref{eq:dyn4}) and (\ref{eq:dyn1})). As blood glucose rises, insulin $I$ is produced to regulate this with a level of efficiency that depends on $\theta_0$ (see (\ref{eq:dyn2}) and (\ref{eq:dyn1})). When glucose goes below the base level $G_b$ then the response $L$ increases glucose with a level of efficiency dependent on $\theta_1$ (see (\ref{eq:dyn3}) and (\ref{eq:dyn1})).

\begin{figure}
    \begin{centering}
\begin{tabular}{c c}
    \includegraphics[width=7cm, height=5.5cm]{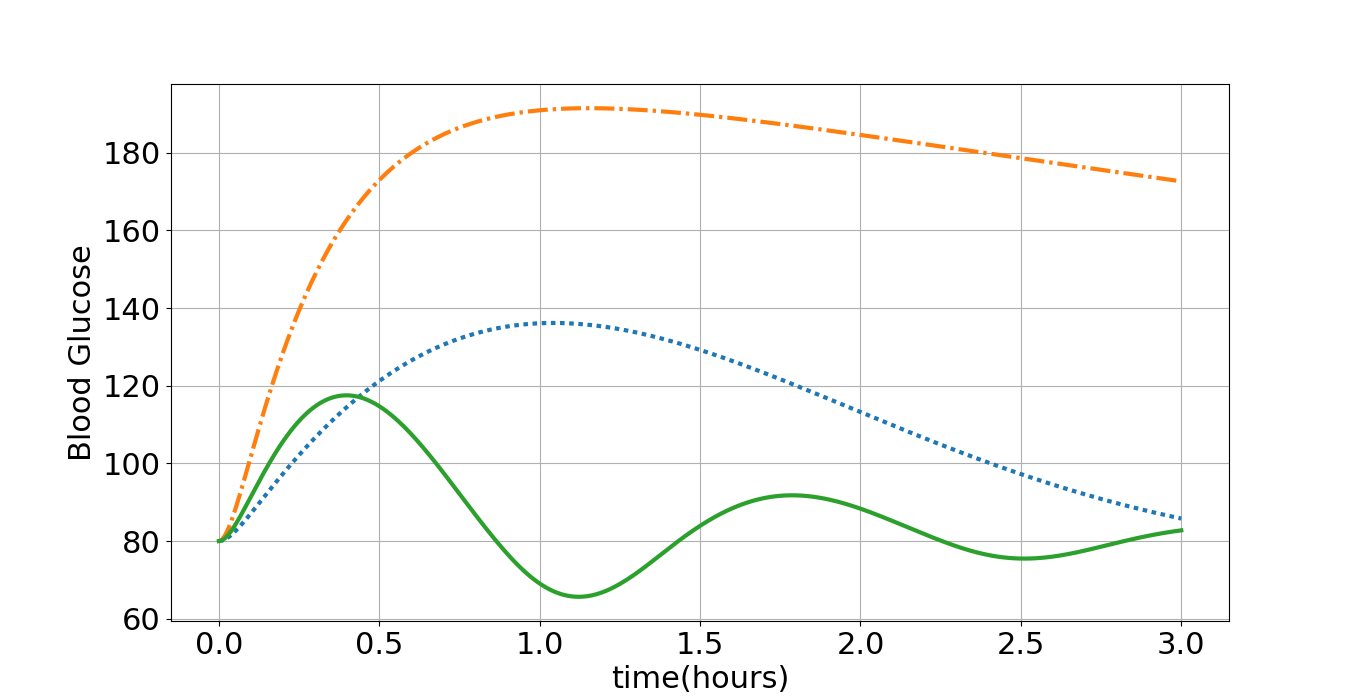} &
	\includegraphics[width=7cm]{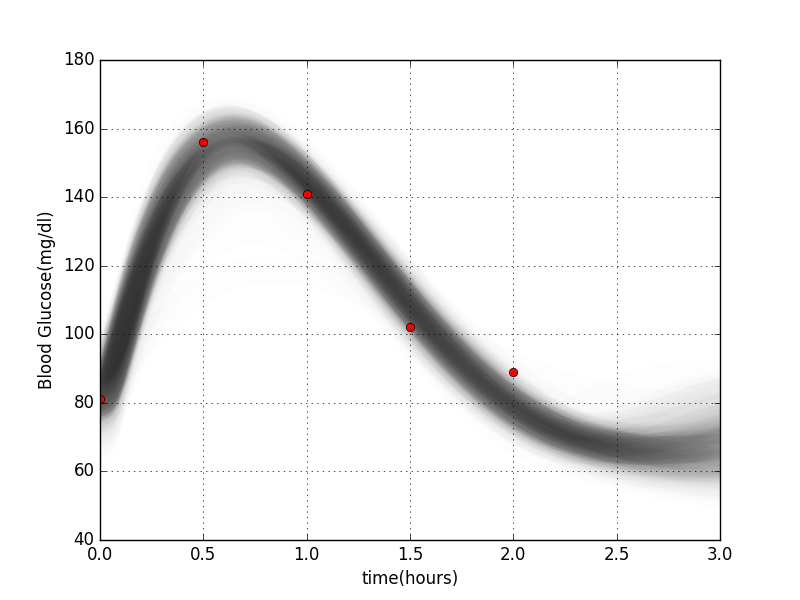} \\
    (a) & (b)
\end{tabular}
\caption{(a) The behavior model $G(t)$ for three different hypothetical patients. The dotted line represents a healthy patient, the broken line is a patient with insulin resistance, and the solid line is an oscillating case where a patient produces an unusually large amount of insulin. (b) Bayesian analysis of OGTT data on a healthy patient. The blood glucose readings are the red dots and the ``shadow'' curve represents the posterior distribution of $G(t)$ (a vertical slice centered on time $t$ would represent a kernel density estimate of the posterior distribution for $G(t)$ at time $t$). Uncertainty in estimating the OGTT curve, modeled with $G(t)$, is depicted here, and predictions of blood glucose are also shown beyond the 2 hour period of the test.}
    \end{centering}
    \label{fig:ogttexamples}
\end{figure}

In figure \ref{fig:ogttexamples} (a) we see how the model behaves with various values of the inference parameters ($G(0)$ is the same for all cases). This displays a healthy patient (dotted line), a patient with insulin resistance (broken line) and a patient who produces an unusually large amount of insulin (solid line). The healthy patient's insulin rises and then returns to normal in 2-3 hours. The potentially diabetic patient does not manage to return to near resting glucose levels after 3 hours, and the patient who produces a large amount of insulin oscillates several times in that period. The last kind of patient is a case which prior OGTT analysis models would have classified as "normal" even though glucose processing does not behave as expected from a typical healthy patient.

\section {Model Inference from venous data}

The data collected in an OGTT test are measurements of $G(t)$ at various times. As described in \cite{odemodel}, the data are modeled with
$$
y_i=G(t_i)+\sigma\epsilon_i;\; \epsilon_i\sim\mathcal{N}(0,1)
$$
where $\sigma$ is set to be 5. The error model of $\mathcal{N}(0,5)$ is taken from clinical opinion of the device's standard error.

For instance, for one actual patient, at times $t=$0:00, 0:30, 1:00, 1:30 and 2:00 hours we obtained glucose measurements of $y= 81,156,141,102, 89~ mg/dl$.  The intent is to use these data to infer the parameters $\theta_0, \theta_1, \theta_2$ and $G(0)$.  This is an "inverse" problem, that is, an inference problem on a regression model in which the regressor is driven by a system of ODEs.

In \cite{odemodel}, the $\mathcal{N}(0,5)$ error model was not subjected to close scrutiny, and accepted mostly on the basis of providing good results. For the present work, AM was able to produce a series of 15 duplicate blood glucose measurements. For these measurements, a venous blood sample was taken, and its glucose content was measured twice (these are not full OGTT results, but simply 15 blood samples). These data are scarce because there is no medical reason to run the same sample twice and the data were collected for this purpose only. The differences between these samples showed that the variance in measurements is actually lower than expected (our estimates show a standard deviation of 3.09 $\frac{mg}{dl}$), but they include two outliers which are more than two standard deviations from zero. These outliers were found regardless of whether the differences were treated as absolute differences or as proportional differences. The amount of data available is still somewhat small, but there is reason to believe that the variance of the error is even smaller than previously expected although the error distribution should be expected to have heavy tails.

For this reason, we use an alteration to the original venous blood model, replacing the Gaussian errors with a scaled t distribution with 4 degrees of freedom. Hence, our new error distribution for venous measurements is
$$
y_i=G(t_i)+\sigma\epsilon_i;\; \epsilon_i\sim t(4)
$$
with $\sigma=2.2$, to make the standard deviation of the error $\sigma\epsilon_i = 3.09$.

\subsection {Priors}

We assign a priori distributions to each of the parameters as follows:

$$
\theta_0\sim Gamma(2,1)
$$
$$
\theta_1\sim Gamma(2,1)
$$
$$
\theta_2\sim Gamma(10,1/20)\; 1{\hskip -2.5 pt}\hbox{I} \{\theta_2 > 0.16\}
$$
$$
G(0)\sim \mathcal{N}(80,100)\; 1{\hskip -2.5 pt}\hbox{I} \{G(0)\in [30,400] \}.
$$
These are selected to reflect estimations of the parameters in real patients. They are chosen to be vague (with regions of high probability extending well beyond usual estimates) so as to avoid overwhelming data even if there are few observations of $G(t)$ and even if the patient has a very unusual glucose profile. The exception is the initial prior for $G(0)$. In this case, patients have been observed who have extremely high levels of fasting glucose. The prior is still appropriately wide simply because any patient with an initial measurement anywhere below 30 or above 400$mg/dL$ is not tested but instead placed into emergency care (a preliminary glucometer blood test is conducted, for removal and immediate tretment of such cases). For additional details on the choice of priors, see \cite{odemodel}.

To perform inference on the model, an MCMC simulation is used to obtain a posterior sample. Calculation of $G(t)$ is done numerically using the LSODA package for ODEs. For the MCMC the t-walk is used \cite{twalk}. Figure \ref{fig:ogttexamples} (b) shows the results of inference on data from a real (healthy) patient. Any vertical grey slice shows a kernel density estimation of the posterior for $G(t)$ at the corresponding time. As we can see, at least with this patient, the model fits the data very well.

\section {Capillary glucose data}\label{glucerror}

While the ODE model mentioned previously works well with standard blood tests, these tests are invasive and bothersome for the patient. They are slow, moderately expensive, and take up valuable time for the laboratory staff. There is, however, an alternative method for measuring blood glucose which is much more practical, called a glucometer. A glucometer is an easily available apparatus which measures blood glucose from capillaries rather than veins. Glucometers were designed for diabeic patients for simple, fast and relatively cheap self measuring of their blood glucose. They are easy to use, yield nearly instantaneous results, and only require a single drop of blood from the patient's fingertip produced with a lancet.  It is usually possible to purchase a glucometer kit inexpensively at a local drugstore. However, the reason that glucometers are not commonly used for OGTT tests is that typically they are not thought to be accurate enough.

While glucometer accuracy is indeed significantly less than venous lab tests, the poor analysis methods commonly used for OGTT analysis may be the reason that they are unusable. The advent of dynamic models for analysis may be a reason to reexamine if it is possible to obtain useful information from glucometer OGTTs.

To this end , AM took data from 65 patients that were visiting the Hospital General in Mexico City for a scheduled regular OGTT test.  Along with the usual OGTT measurements, these patients also had their blood tested with a glucometer at the same time as the blood sample used for the normal tests. The glucometer measurements are - essentially - a duplicate OGTT for the same patient, measured with a glucometer and capillary blood. In order to ensure the patient's safety, AM also habitually tests venous blood with a glucometer before giving patients the glucose concentrate (this only takes a few seconds and ensures that if a patient's glucose is too high then he/she can get immediate treatment). This is convenient because it provides a measurement which differentiates the effect of the blood (venous vs capillary) from the effect of the glucometer. 

To adapt to data of poorer quality, we adjust our error model for measurements. Literature about glucometer error indicates that the conditions greatly affect the reliability of glucometer measurements. OGTT conditions are quite singular in that they are performed by a professional in a carefully controlled environment, but with a patient whose body is undergoing a very specific -and somewhat unusual- kind of stress.  

\begin{figure} 
	\centering
    \includegraphics[scale=0.5]{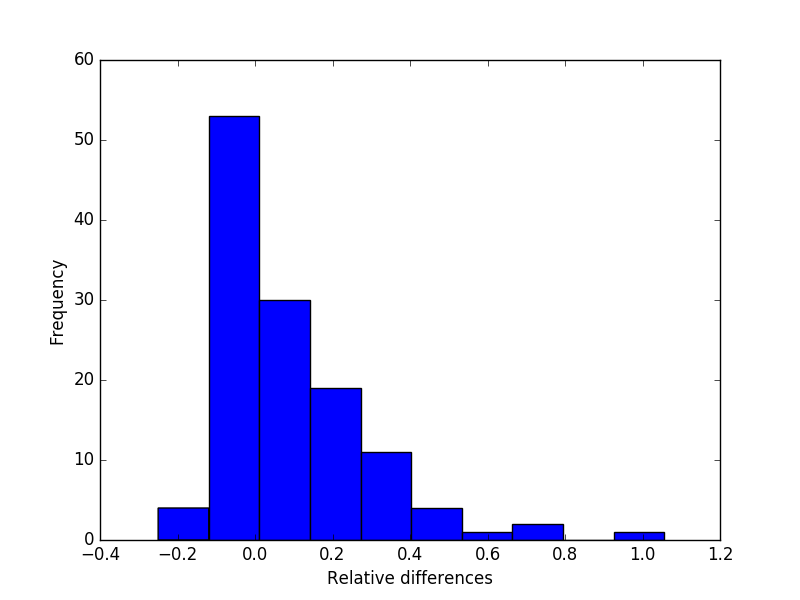}
    \caption{Histogram of the relative differences between glucometer capillary and  lab venous blood glucose measurements in AM's dataset.}
    \label{fig:reldifs}
\end{figure}

From AM's dataset, the proportional differences between glucometer and venous blood glucose measurements are charted in a histogram, seen in figure \ref{fig:reldifs}.

One important detail to note is that along with the wide spread of values, there is also a clearly visible bias. This bias makes medical sense, since blood in the capillaries delivers glucose to muscles before entering the veins. The errors of the venous test are negligible in comparison to the errors of the glucometer test, so modelling the difference between the glucometer and the venous data is tantamount to modelling glucometer error.

A model was selected which gives the error as the sum of a gamma distributed bias due to the loss of glucose in the capillaries, and a normally distributed measurement error. Since the bias is driven by biological processes in the patient's body, it is assumed to be equal for all measurements across a single OGTT, but different from patient to patient.

Point estimates are obtained for the bias and error parameters from the data and we obtain a model for capillary data. This model is now
$$
y_i=G(t_i)+b+\epsilon_i;\; b\sim Gamma(0.4,4.3)\; \epsilon_i\sim\mathcal{N}(0,100) .
$$

\begin{figure}
    \begin{centering}
        \includegraphics[scale=0.8]{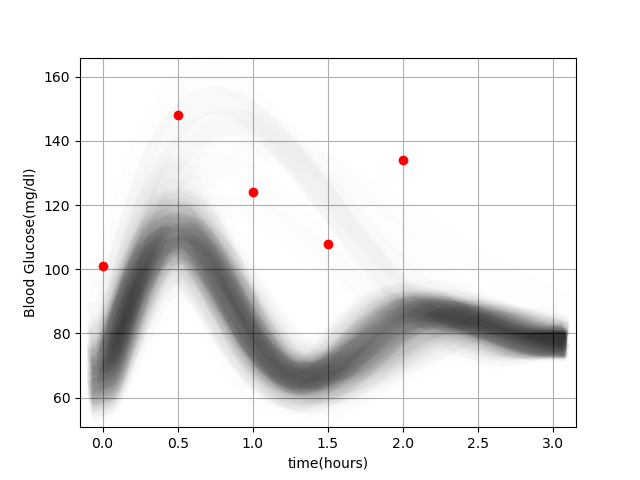}
        \caption{Inference on a patient using capillary data. In contrast to venous data, glucometer measurements are known to be biased, and our error model takes this into account. As a result, the OGTT curve estimated from capillary data does not necessarily closely approximate the data. From this curve alone, it can be very difficult to tell at a glance how well the model describes the data. One way to do so is to compare it to a venous OGTT from the same patient.}
        \label{capogtt}
    \end{centering}
\end{figure}

Figure \ref{capogtt} shows a single patient's capillary OGTT from glucometer measurements. In contrast to figure \ref{fig:ogttexamples} we observe that the path of the estimated curves does \textit{not} necessarily pass close to the data. This is because the data is known to be biased, and inference takes this into account by using the above error model. Merely by looking at the graph and the data, it is difficult to know whether the estimation is accurate, and hence it must be compared to a venous OGTT from the same patient.

\section{Results}

\begin{figure}
    \begin{tabular}{cc}
        \includegraphics[width=7cm, height=5.5cm]{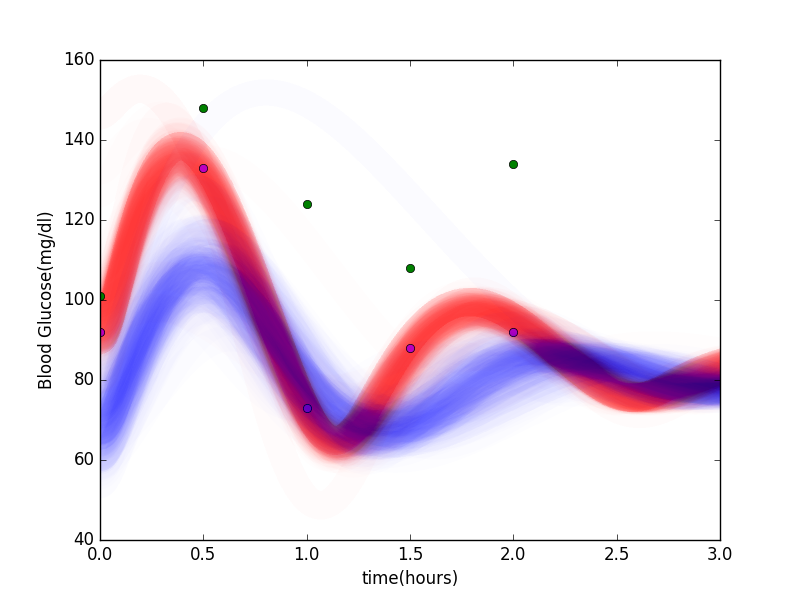} &  \includegraphics[width=7cm, height=5.5cm]{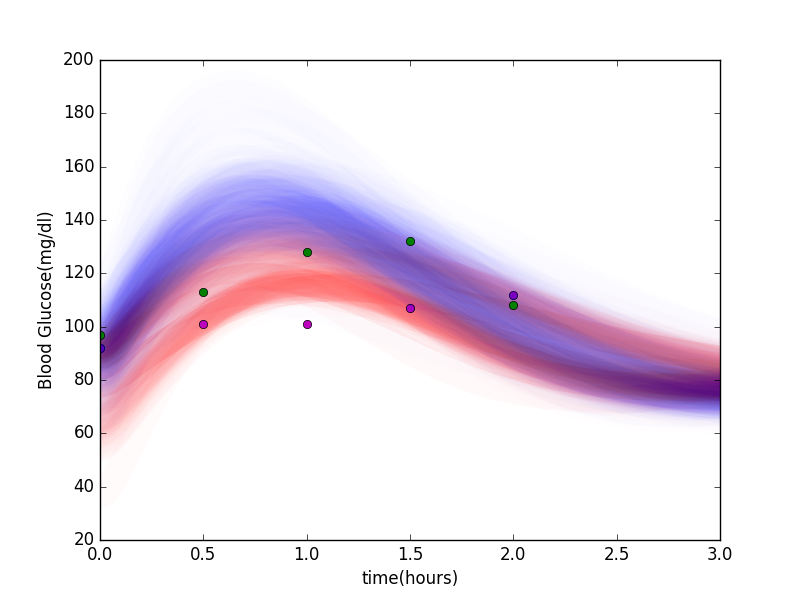} \\
        (a) & (b)\\
        \includegraphics[width=7cm, height=5.5cm]{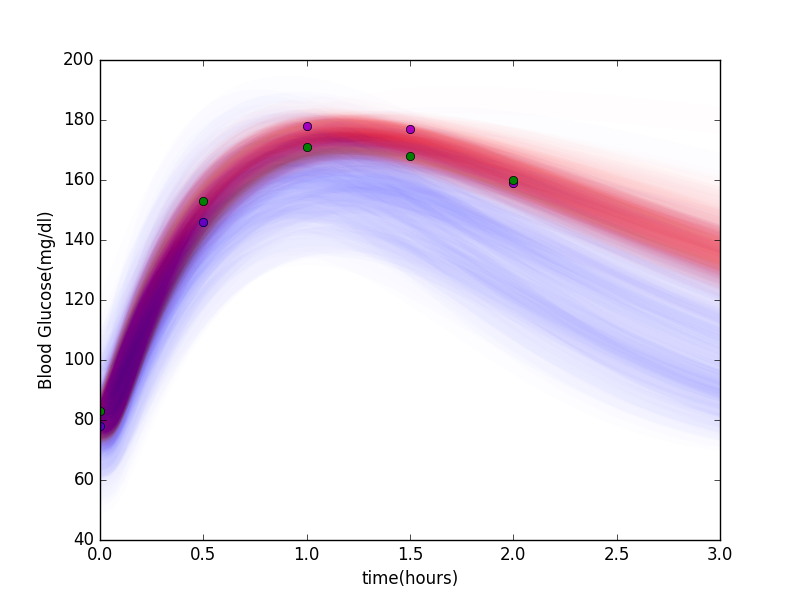}
        &
        \includegraphics[width=7cm, height=5.5cm]{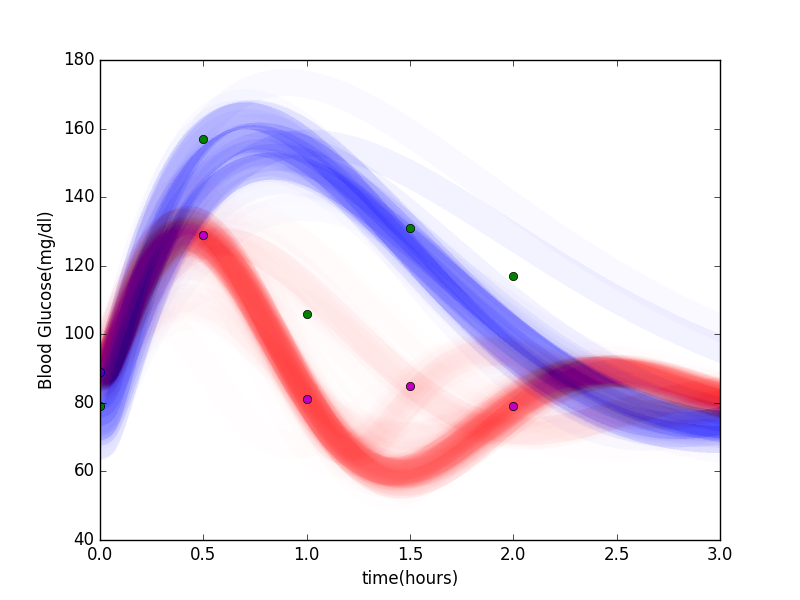}\\
        (c) & (d)
    \end{tabular}
		
                \caption{Data from four real patients. Using venous data we get the magenta dots and the red curves. Using capillary data we get the green dots and the blue curves. In (a) we observe that the two sources of blood produce discordant  data, but the error model for capillary blood is able to adjust for this and we achieve somewhat similar estimations of $G(t)$. This is the same patient from figure \ref{capogtt}. For (b) we observe a patient whose capillary data is similar to the venous data. The bias is minimal, and the model is flexible enough to account for this situation as well, resulting in a capillary OGTT that is also very similar to the venous OGTT. For (c) we have the most common sort of issue that is seen with OGtts from capillary data. In this patient's case, the venous and capillary OGTT's match rather well, but the capillary model has too much uncertainty to derive any clear conclusions about the patient's condition. For (d) we have a rare case where the capillary OGTT and the venous OGTT simply do not match. For the data from this patient, the model has produced an error which may be unacceptable. \label{fig:good6}}
\end{figure}

The model was used to perform inference from capillary and venous blood data from 65 patients. The results for two such patients can be seen in figures \ref{fig:good6}(a) and (b). The first patient is a case where venous and capillary data differ greatly. Nonetheless, the capillary blood error model is able to compensate for the heavily biased data and produces a posterior estimate which behaves similarly to the venous data estimate. The second patient is a case where venous data is similar to the capillary data. The observed bias is minimal, and yet the model is flexible enough to understand that compensation is not required in this case, producing an estimate which corresponds quite closely with estimates obtained from venous blood. 

Despite the advantages of our model to fit capillary glucose readings, we have detected some cases where it still does not quite approximate the same results as venous OGTT tests. There are two main problems that can occur. The first consists of cases where we see a venous and capillary estimate of $G(t)$ for the same patient, but although the capillary test produces an estimate which overlaps the estimate from the venous test, the estimate of $G(t)$ using the capillary data has such excessive variance that it would be very difficult to draw any conclusions regarding what it indicates about the patient's glucose regulating condition. Note, however, that this may not represent a significant problem. If a patient's estimate of $G(t)$ from a glucometer data has excessive variance, then the capillary OGTT may be called inconclusive and a conventional venous test may be recommended for the patient. The supplementary material contains images of posterior estimations for all patients. The behavior just explained may be seen in patients 1, 3, 18, 19, 36, 39, 54, 55, 56, and 58, that is 15.4\% of the total number of patients tested. An example of this situation (patient 55) is seen in figure \ref{fig:good6}(c).

The second issue is the case where estimates using capillary glucometer data simply do not match the estimates arising from venous data. Sometimes this is apparent from visual examination of the data alone, since the data from the capillary OGTT and venous OGTT exhibit different behavior and no adjustment to the error model (nor, in all likelihood, to the ODE system) will solve this issue. From 65 patients for whom this test was performed, only 3 of them produced this sort of unacceptable error. These cases are patient 4, patient 31, and patient 51. An example of this situation (patient 31) is seen in figure \ref{fig:good6}(d).

\section{Conclusions}

The new capillary model has produced unacceptable results in about 5\% of the tested patients. While this does seem like an alarmingly high rate of error, it must be compared with the existing base error rate.
While data on error rates in diagnosis from classical OGTT analysis (without the ODE model) are not readily available, there is reason to suspect that errors may well occur even more often than with tests using the capillary model (\cite{ogttcriteria} reports misdiagnosis rates for some versions of classical diagnostic criteria).

The practical benefits in using capillary blood and glucometers are significant: Not only are venous tests expensive and bothersome, but for many people they may simply be unavailable because of lack of access to a medical facility with the proper equipment. Glucometers, on the other hand, are frequently sold in pharmacies, intended for home use, and are relatively inexpensive. Given an appropriate glucose syrup, a glucometer, and appropriate analytical tools to infer from capillary tests (like those here suggested) it would be theoretically possible to perform an OGTT tests cheaply in a basic facility by a nurse. Note that an user friendly software may be easily coded for analysis of the glucometer readings, to produce a plot as in figure \ref{capogtt} and draw all necessary conclusions from the test.

It is worth noting that numerous errors are reported which originate from stages other than the statistical analysis, and that in at least one case, it was precisely the capillary OGTT which allowed the detection of a problem in the venous OGTT.  Namely, upon first examination, four unacceptable cases were found, but the raw data for these cases was reviewed. After reviewing the raw data, one case was found to be the result of an error when filing the output of the venous test. The initial clinical treatment for this patient was incorrect, and without this careful study of the capillary data, the error would not have been detected. When the correct data were found, we repeated our analysis and found that the venous and capillary data matched well.

If it is practical to obtain a proper venous OGTT test, this is obviously preferred. However, we believe that a capillary OGTT is an alternative which should be considered for situations when venous OGTT tests are impractical.

\section{Data availability}
The data that support the findings of this study are available on request from the corresponding author.

\medskip
\bibliography{vencap}
\end{document}